\documentclass[twocolumn,showpacs,preprintnumbers,amsmath,amssymb]{revtex4}
\usepackage{graphicx}
\usepackage{epsfig}
\usepackage{dcolumn}
\usepackage{bm}
\usepackage{color}

\begin{document}

\preprint{}
\title{Cluster size effects in the magnetic properties of Fe$_p$-Al$_{q=1-p}$ alloys}
\author{J. B. Santos-Filho$^{1}$}\author{A. V. Santos S\'a$^{1}$}\author{T. S. de Araujo Batista$^{2}$} \author{J. A. Plascak$^{3,4,5}$}
\affiliation{$^1$Instituto Federal de Ci\^encias e Tecnologia de Sergipe 49100-000  S\~ao Crist\'ov\~ao, {\it Sergipe}, Brazil}
\affiliation{$^2$Instituto Federal de Ci\^encias e Tecnologia de Sergipe 49060-108 Aracaju, {\it Sergipe}, Brazil}
\affiliation{$^3$Departamento de F\'isica, Universidade Federal de Minas Gerais, 
C. P. 702, 30123-970, Belo Horizonte, MG, Brazil}
\affiliation{$^4$Departamento de F\'isica, Centro de Ci\^encias Exatas e
da Natureza, CCEN, Universidade Federal da Para\'iba, Cidade Universit\'aria,
58051-970 Jo\~ao  Pessoa, PB - Brazil}
\affiliation{$^5$Department of Physics and Astronomy, University of Georgia, 30602 Athens GA - USA}

\date{\today}

\begin{abstract}
A spin-1/2 Ising model, defined in the body centered cubic lattice, is used to describe some 
of the thermodynamic properties of Fe$_p$-Al$_q$ alloys, with $p+q=1$. The model assumes,
besides the nearest-neighbor exchange coupling, the existence of further next-nearest-neighbor 
superexchange interactions, where the latter ones depend on the aluminum atoms cluster size. 
The Ising system so considered is studied by employing Monte Carlo simulations, using a 
hybrid algorithm consisting of one single-spin Metropolis move together with one 
single-cluster Wolff algorithm allied, in addition, with 
single histograms procedures and finite-size scaling techniques. Quite good fits to the 
experimental results of the ordering critical temperature, as a function of Al concentration in the 
range $0\le q<0.7$, are obtained and compared to more recent theoretical approaches done on the
same alloys. 
\end{abstract}

\maketitle


\section{Introduction}
The magnetic and structural properties of iron aluminides have been widely studied for more than 
a century. This interest is certainly due to their possible industrial application properties 
such as oxidation and corrosion resistance, good ductility at room temperature, relatively low 
density, high magnetic permeability  value, good vibration damping and insulation properties 
(for an up to date review of mechanical and some magnetic properties on such compounds see, for 
instance, reference  \cite{Palm2019} and references therein). Recent studies have also shown 
that Fe-Al alloys are promising candidates for data storage, due to the presence of some magnetic 
patterns appearing at room temperature  \cite{Sort2006, Menendez2009, Bali2014}. At the same time, the 
magnetic properties of Fe-Al alloys are strongly linked to their chemical structure, since 
depending on composition and synthesis method, they can behave either as a ferromagnet or a
paramagnet at room temperature  \cite{Besnus1975, Oubelkacem2010, Rodriguez2016}. It is also 
possible to induce in these compounds new types of cooperative phenomena, such as superparamagnetism 
and reentrant spin glass like phases  \cite{Arrott1959,Shull1959,Boni1986,Takahashi1996}. 
These characteristics have surely made this alloy one of the most studied binary systems 
in both experimental and theoretical research points of view.
	
In particular, the disordered Fe-Al alloys, known as $A2$ or $\alpha$-Fe phase, present a 
lattice structure, with compositional disorder, in the body centered cubic (bcc) lattice 
 \cite{Zamora2009, Trautvetter2011}.  They may be represented by Fe$_{p}$Al$_{q}$, where 
$p$ stands for the concentration of iron atoms and $q$ for the concentration of aluminum 
atoms, with $p+q=1$. In this way, $q = 0$ corresponds to the pure iron case, which has a 
ferromagnetic to paramagnetic critical phase transition at a temperature $T_c=1040$ Kelvin. 
Due to the fact that the Al atoms have no magnetic moments, they act indeed as a diluter, 
and the transition temperature $T_c(q)$ should depend on their concentration $q$. Nevertheless, 
the corresponding experimental phase diagram of these alloys in the range of $0 <q <0.2$ shows 
a quite small variation of the critical temperature $T_c(q)$ as a function of the concentration 
$q$. Such unexpected result is in clear contrast to the theoretical predictions, where a finite 
negative slope $dT_c(q)/dq|_{q\rightarrow 0}<0$ is obtained for the transition line  
\cite{Domb1983}. For $0.2<  q <0.3$, on the 
other hand, the  critical temperature shows a steep slope and eventually vanishes at  $q \sim 0.7$.
	
In order to try to explain this rather anomalous behavior of the phase diagram, several models 
(including classical and quantum) and theoretical procedures (including analytical approximations 
and computer simulations), have been proposed in the literature (without any intention of making 
an extensive review of the previous studies in these alloys, we can cite, for instance, references 
 \cite{Perez1986,Plascak2000,sala,mota,osorio,Contreras2005,aguirre,lara,dias1,dias2,Diaz2012,
Freitas2013,freitas}). It should be mentioned that, despite localized spin Hamiltonians are not completely 
appropriate for the study of these metallic alloys, there are some experimental evidences and 
first principles calculations making possible the choice of  localized models instead of the 
itinerant ones, as discussed, respectively, in Refs.  \cite{Perez1986} and  \cite{Ghosh2015}. 

In all the theoretical treatments in the above references, the common ingredient is the 
assumption that the exchange interaction is dependent on the Al concentration. This is due to 
the observed expansion of the lattice parameter that is produced when the larger Al atoms 
replace Fe atoms in the bcc network  \cite{Perez1986}. The second most common feature present 
in almost all these approaches is the presence of a superexchange interaction induced by the 
Al atoms  \cite{Plascak2000}. This extra ferromagnetic interaction thus links next-nearest-neighbor 
Fe atoms in the lattice and has also been assumed to depend on Al concentration but be independent 
of the number of Al 
atoms that are clustered together. Indeed, with this assumption, a better agreement with the 
experimental results of the magnetic phase diagram has been obtained. It has been noticed that 
the results are qualitatively independent whether the model is classical or quantum 
 \cite{sala,Contreras2005}, or even being treated according to some approximation scheme based 
on mean-field like procedures in the so-called pair approximation. Effective field approaches have also 
been used to treat the theoretical model \cite{dias1,Freitas2013,freitas}, as well as Monte Carlo 
simulations  \cite{aguirre}. Although most 
of the models employed spin-1/2, some spin-1  \cite{dias2,freitas} and spin-2  \cite{Freitas2013} 
models have also been considered. In this case, the best results so far for the phase diagram 
have been achieved by Freitas et al. by using an effective field theory within the one-spin 
cluster approach \cite{Freitas2013}. 

It is worthwhile to mention, however, that the effective field theory in Ref. \cite{Freitas2013} 
considers an Ising system with spin-2 and only nearest-neighbor exchange couplings, without any 
induced superexchange interaction between Fe atoms. In addition, this nearest-neighbor exchange 
coupling exponentially depends on the Al concentration $q$. With these assumptions, the best fit to the experimental data has
led to an Fe-Fe interaction that increases (of the order of 30\%) as the Al concentration increases (up to 
$q\sim 0.22$), and then decreases to almost zero for $q\sim 1$. This counter-intuitive increase in 
the Fe-Fe interaction has been recently reported on these compounds, by using first principles 
calculations, for alloys with rather high Al concentrations, $0.35 \le q \le 0.5$, and out of 
the anomalous region  \cite{Ghosh2015}. This increase of the nearest-neighbor exchange coupling 
with $q$ has been ascribed, in this range of concentrations, to the fact that as the 
concentration of Al increases, the picture becomes closer to Fe clusters immersed in a sea of Al, 
with such clusters of now {\it magnetic} impurities having higher exchange energy and larger 
moment as compared to greater Fe compositions  \cite{Ghosh2015}. However, we do not expect this 
feature to happen in the anomalous region, since in this region the Al concentration is still 
small enough to likely behave as a sea for the magnetic atoms. The more plausible role seems indeed to 
have the Al clusters immersed in a sea of Fe atoms instead. In addition, as will be discussed below, 
Al clusters of different sizes should behave in different ways. 

Another important point concerns the value of the Fe spin. As has been reported in Ref. \cite{Ghosh2015}, the spin value of 1/2 seems more appropriate for this system than the value 2 considered in Ref. \cite{Freitas2013}, and even the value 1 in Ref. \cite{dias2}. Recall still that the fitted crystal field parameter (in units of the ferromagnetic exchange interaction) of the spin-1 Blume-Capel used for these alloys \cite{dias2} is of the order -10. This rather large negative crystal field means that the probability of occupancy of the zero states of the spin are quite small, leaving only the $\pm 1$ spin states more probable to occur.   

We can now draw our attention to the question of the physical effects of the cluster sizes. One step towards the analysis of the behavior of small clusters of atoms has been achieved from 
the advent of nanotechnology. It is now known that the elements behave quite differently when 
their dimensions are reduced to the point of being formed by just a ``countable'' number of atoms  
\cite{Koch2006}. Recently, researchers have shown, for example, that a cluster of only $102$ gold 
atoms behaves as a giant molecule, instead as of a metal. However, clusters with $144$ atoms 
exhibit in fact a metallic behavior  \cite{Mustalahti2015}. Since we must visualize the Fe-Al 
system as being composed of different aluminum clusters dispersed in an iron lattice, we can,
by analogy, also assume that the number of Al atoms in a given cluster inside the alloy can, 
in some way, differently influence the induced superexchange interaction. For instance, only 
clusters up to a given size are able to switch on the superexchange interaction, while larger 
clusters are not. This comes to be a reasonable assumption because depending on the superexchange 
strength, the theoretical slope of the critical temperature as a function of Al concentration 
turns out to become positive in the limit $q\rightarrow 0$, which is indeed contrary to what has
been experimentally observed.

The main goal of this study is exactly to take into account the Al cluster size dependence on 
the induced superexchange interaction, and treating the regular nearest-neighbor exchange 
interaction in the same way as done in previous works. For this purpose, we will consider an 
Ising model and will employ Monte Carlo simulations with a hybrid algorithm consisting of 
single-spin flip Metroplis and single-Wolff cluster algorithms, together with histograms 
techniques and finite-size-scaling analysis. Although simulations are, in general, not suitable 
for fitting experimental results, here we have additionally resorted to a mixed Fortran-Python 
language subroutines to an easier treatment of the computer simulation data. At the same time, Monte Carlo simulations are in fact very well suited to microscopically study the effects of different Al cluster sizes present in the system for any given aluminum concentration. 

The plan of the paper is the following. The model and some details of the simulations are 
presented in the next section. The results are presented in Section \ref{results} and some 
concluding remarks are addressed in the last section.

\section{Model and simulations}
\label{model}

\subsection{Model}

As we will assume that aluminum atoms have no magnetic moment, the system under investigation can be well represented 
by a quenched site-diluted Ising model with Al as dilutors. The Hamiltonian for such spin system can then be written as
\begin{eqnarray}\label{hamil}
\begin{split}
\mathcal{H} =&-J_1(q)\sum_{\langle nn\rangle} \epsilon_i \epsilon_jS_{i}S_{j} \\ 
 & -J_{2}\sum_{\langle nnn \rangle}\varsigma_{ij}  \epsilon_i \epsilon_j S_{i}S_{j}~,
\end{split}
\end{eqnarray}
where $S_i$ represents an Ising spin-$1/2$ 
at the sites of a body centered cubic (bcc) lattice, meaning that we can consider $S_i=\pm 1$. 
The first sum is over nearest-neighbors $\langle nn\rangle$ pairs and the second sum over 
next-nearest-neighbors $\langle nnn\rangle$ pairs.   $J_1(q)$ is the nearest-neighbor interaction 
given by  
\begin{equation}
J_1(q)= J(1 - Aq)~,
\label{eq:j}
\end{equation}
where $A>0$ is an additional theoretical parameter that simulates the decrease of the exchange 
interaction due to the increase of the lattice parameter as the larger Al atoms concentration 
$q$ increases. $J$ is the Fe-Fe exchange interaction in the pure undiluted system. The next-nearest-neighbor superexchange interaction $J_2$ induced by the Al atoms, is given by
\begin{equation}
J_2= BJ~,
\label{eq:j2}
\end{equation}
where the parameter $B$ simply gives the strength of $J_2$ in units of $J$. Note that, with the above definition, 
$J_2$ is here formally independent on the Al concentration $q$, contrary as has been 
assumed in previous works. 
In Eq. (\ref{hamil}), $\epsilon_i$ are quenched, uncorrelated random variables, representing 
the existence of two kinds of particles in the system, namely the magnetic ones (Fe) with 
$\epsilon_i=1$, and non-magnetic ones (Al) with $\epsilon_i=0$.  The variable $\epsilon_i$ 
is chosen according to the bimodal probability distribution
\begin{equation}
P(\epsilon_i) = p\delta(\epsilon_i-1) + q\delta(\epsilon_i).
\label{eq:p}
\end{equation}

The extra variables  $\varsigma_{ij}$ in the second sum of Eq.(\ref{hamil}) represent the existence 
of superexchange interaction between sites $i$ and $j$. However, this new interaction between 
iron atoms is now induced only by aluminum atoms on small clusters. The cluster size $\zeta$ 
that induces superexchange interaction between the next-nearest-neighbors is thus another 
adjustable parameter of the model and is given by just the number of Al atoms, independent of the 
cluster shape. In this way, $\varsigma_{ij}=1 $ for next-nearest-neighbors $i$ and $j$ 
adjacent to the border of the cluster, if the cluster size is smaller than $\zeta$, and 
$\varsigma_{ij}=0 $ otherwise. Fig. \ref{fig0} depicts a situation where $\zeta= 4$ 
on a two-dimensional lattice. 
 In Fig. \ref{fig0}(a) the tree Al atom cluster is smaller than $\zeta= 4$, thus being able to generate the next-nearest-neighbor interactions $J_2$, while in Fig. 
\ref{fig0}(b) a cluster of five Al atoms does not induce any $J_2$. Of course, all clusters
with four spins also induce the superexchange interaction.
Note that, with this assumption, 
although $J_2$ does not explicitly depend on $q$, it will be influenced 
by the Al concentration. The reason is the following. In a quenched dilution, the increase in concentration 
$q$ leads to the formation of larger Al clusters. Since the probability of finding a cluster 
with size $\zeta$  depends of $q$, the parameter $\zeta$ will indirectly connect the effective number of
induced $J_2$ interaction to the concentration $q$. 

\begin{figure}[]
  \centering
   \hspace{0.8cm}
 {\scriptsize\textbf{(a)}}\\
   \includegraphics[height=4.5cm]{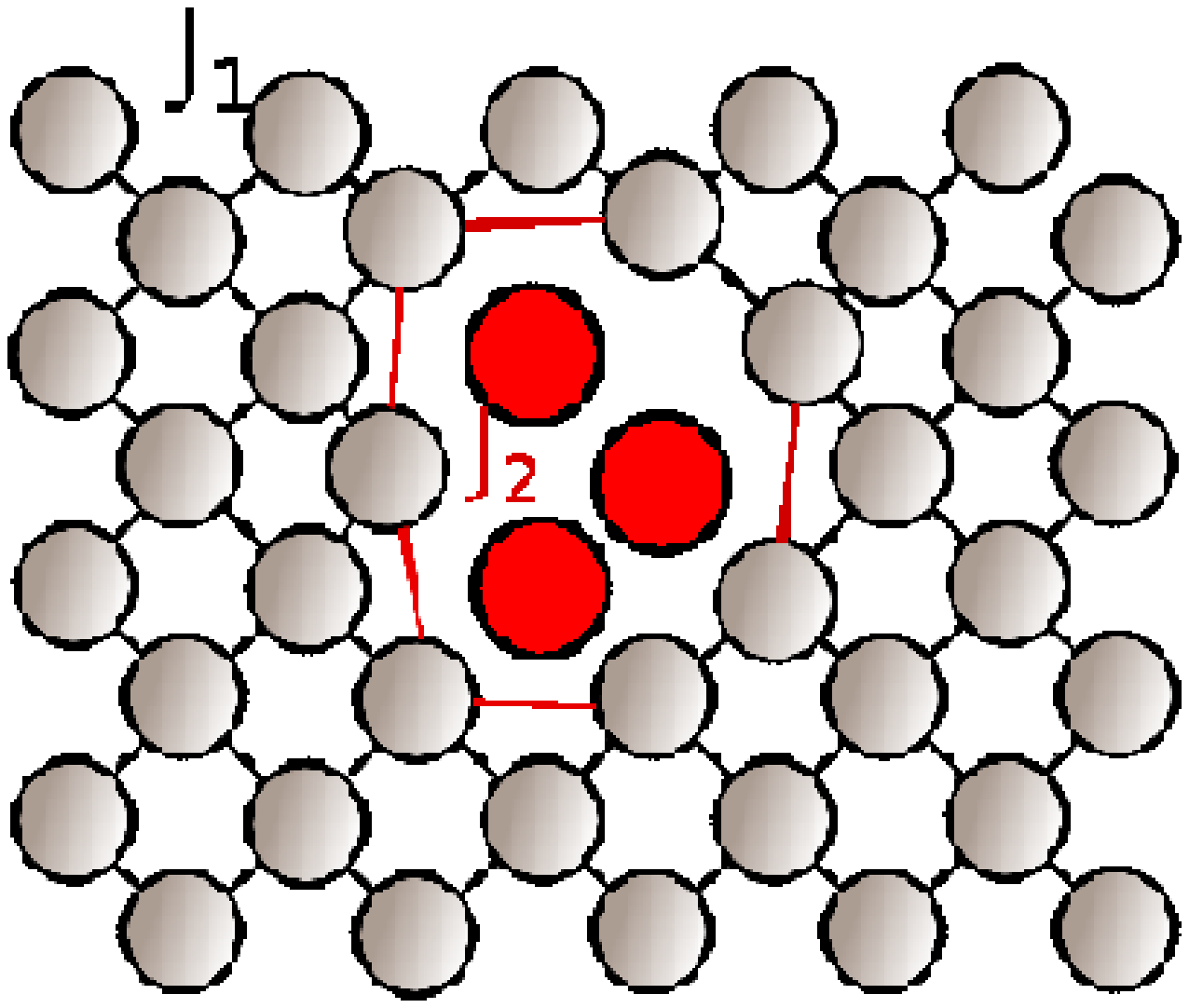}
\end{figure}
\begin{figure}[]
   \centering
   \hspace{0.8cm}
 {\scriptsize\textbf{(b)}}\\
   \includegraphics[height=4.5cm]{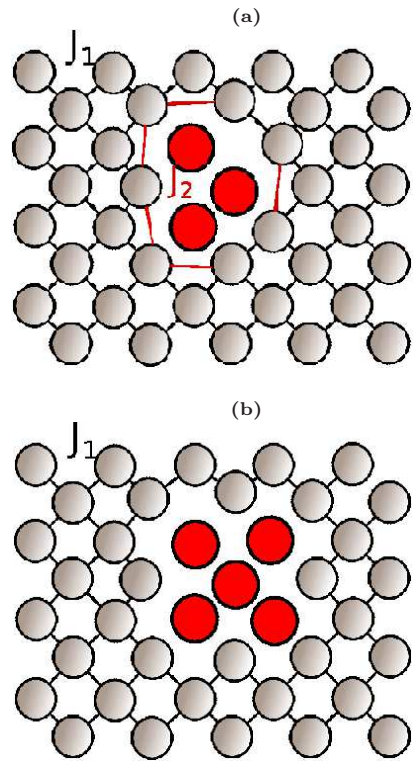}
\caption{(Color online) A two-dimensional lattice sketch of the induced superexchange 
 interaction $J_2$ by 
 the aluminum cluster for $\zeta=4$. Connected circles are Fe atoms linked by the exchange 
 interaction $J_1=J_1(q)$, and larger disconnected circles are Al atoms. In (a) the tree 
 Al atoms are able to generate the next-nearest-neighbor interactions $J_2$, while in (b) 
 a cluster of five Al atoms does not induce any $J_2$.}
\label{fig0}
\end{figure}

In summary, the present model for these alloys has in fact just three adjustable parameters: parameter $A$ that quantifies the variations of the exchange interaction with the lattice constant; parameter $B$ that quantifies the intensity of the induced superexchange interaction; and parameter $\zeta$ that limits the superexchange interaction to small Al cluster. The parameter $J$ is obtained from the critical temperature $T_c$ ($=1040$ Kelvin) for the pure Fe compound ($q=0$) or, preferably, one can simply construct the reduced temperature phase diagram by plotting the non-dimensional quantity $T_c(q)/T_c(0)$, as a function of $q$, in order to easier obtain the fitted parameters. 

\subsection{Simulations}

The above Hamiltonian has been studied through Monte Carlo simulations. First, we prepare a diluted sample, i.e., for a given value of $q$, we build a randomly diluted finite lattice of linear dimension $L$. Next, we locate and measure all the Al cluster sizes. For all cluster sizes having a number of diluted sites smaller than $\zeta$, the superexchange interaction is switched on between the surrounding next-nearest-neighbor magnetic atoms. Then, we use a hybrid Monte Carlo simulation consisting of one single-spin flip Metropolis algorithm  \cite{Metropolis, newman1999monte}, and one single-cluster Wolff algorithm  \cite{wolff1989collective}. This hybrid Monte Carlo method  \cite{creutz, pawig} has shown to reduce correlations between successive configurations in the simulation. In the present case, one MCS per spin consists of one Metropolis sweep followed by one single-cluster Wolff algorithm (of course different choices of Metropolis and Wolff runs can be implemented, however, the present one turned out to be more efficient if one considers correlations between configurations and the available computer time).  
  
We have computed the magnetization, magnetic susceptibility, and the fourth-order Binder cumulant given, respectively, by 
\begin{eqnarray}
m&=&\frac{1}{N}\sum_{i=1}^{N}S_{i},\\
\label{mag}
\chi &=& N \frac{\langle m^{2}\rangle -\langle m \rangle 
^{2}}{T},\\
U_{4}&=&1-\frac{\langle m^{4}\rangle}{3\langle m^{2}\rangle^2},
\label{U1}
\end{eqnarray}
where $N=2L^3$ is the total number of lattice sites. The lattice sizes raged from $L=5$, $10$, $15$, $20$, $25$ and $30$. The lattice size $L$ and concentration $q$ values have been chosen so that $q \times N$ gives an integer number. All the above values of $L$ have only been used for $q = 0$, $q = 0.3$ and $q = 0.7$, while for the other points in the phase diagram we have just used $L = 10$ and $L = 20$. It should be stressed that the use of only two lattice sizes can be done, in this case, because the scale of the reduced temperature in the phase diagram does not require a great precision, and also because the error in the simulation results coming from these two lattices alone, when compared to the finite-size extrapolation with larger systems, are smaller than the corresponding symbol sizes used for the experimental and simulational data. 

Initially, the simulation runs comprised $ 1 \times 10^4$ MCS per spin for equilibration and  the measurements were made over more $1 \times 10^4$ MCS at different temperatures, in order to locate the temperature at which the maximum susceptibility occurs or the cumulants cross. Then, a new simulation is performed at this temperature with $ 3 \times 10^4$ MCS per spin for equilibration and new measurements of the corresponding thermodynamic quantities have been made by using single histogram techniques \cite{ Ferrenberg1988, Ferrenberg1991} with $1 \times 10^5$ additional configurations. For each value of the concentration $q$, this whole process is repeated by constructing $50$ different randomly diluted samples. 

\section{Results}
\label{results}

\begin{figure}[htb]
\begin{center}
\includegraphics[width = 8.5cm]{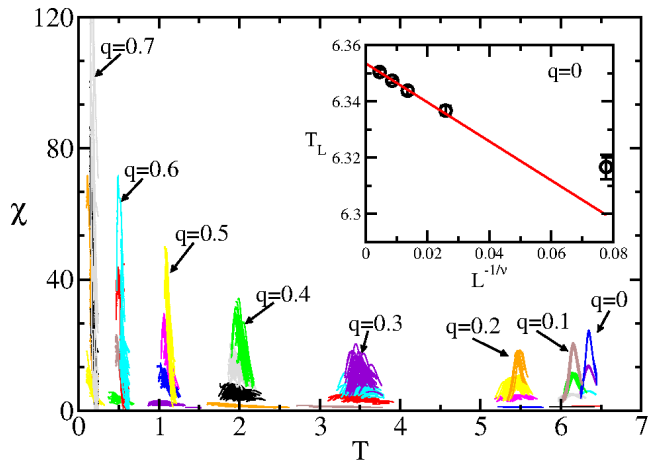}
\end{center}
\caption{(color online) Magnetic susceptibility as a function of the temperature for different samples and concentration $q$ values. In this example, for $L=5$ to  $L=20$; $\zeta=60$, $A=1.25$ and $B= 0.75$. For a question of clarity, the individual Monte Carlo data have been omitted and only the lines are shown in this figure. The inset shows the size dependence of the finite-lattice effective critical temperatures estimated from the maximum of $\chi$ for $q=0$, and in this case using
$L = 5$ to $30$, with $\nu =0.6302$ \cite{Butera2000}. The value found  is $Tc =6.3535(3)$ without considering the smallest lattice.} 
\label{fig:1}
\end{figure}

In Fig. \ref{fig:1} we have typical results  of the magnetic susceptibility $\chi$ as a function of temperature $T$, for each of the 50 simulated samples, for concentrations ranging from $q=0$ to $q=0.7$ with steps of $0.1$, lattice sizes $L=5,~10,~15$ and $20$, $A = 1.25$, $B= 0.75$, and $\zeta = 60$. The temperature here is measured in units of $J/k_B$, where $k_B$ is the Boltzmann constant. We have also obtained results for other values of the theoretical parameters $A$, $B$ and $\zeta$. In Fig. \ref{fig:1} we have chosen the ones that best fit the experimental data. It is clear from this figure that the temperature of the susceptibility peak decreases as the dilution increases, as expected, and above $q=0.7$ the transition temperature should vanish (the site diluted critical threshold for the bcc lattice is $q_c=0.7540385(10)$  \cite{ziff}).  We can also observe that the susceptibility peak becomes more dispersed as the concentration of Al atoms increases from $q =0$ to $q = 0.4$. For example, for $q=0.1$, in the scale of that figure, the susceptibility curves do not show a sensitive dispersion, and are quite close together. This is a result of the smaller sizes of the Al clusters. On the other hand, for $q = 0.3$ and $q=0.4$, the peaks show a large variation among the samples, due now to the different cluster sizes present in the random system. As the concentration of Al atoms further increases, there is a kind of saturation of the cluster sizes inside the sample and the dispersion turns out to be again less pronounced. 

The effective transition temperature $T_L(q)$ is thus obtained from the average temperature of each of these peaks with the standard deviation being the error estimate. As the histogram for each sample has been computed at different temperatures, the estimate of $T_L(q)$ and its error are statistically obtained from 50 different peaks, instead of locating the peak of the averaged susceptibility. The results from both approaches should be, however, equivalent \cite{ivan}. As an example, the inset in Fig.\ref {fig:1} shows the size dependence of the effective critical temperatures for $q=0$ and $L = 5$ to $30$ with the critical exponent $\nu =0.6302(4)$ taken from the Ref. \cite{Butera2000}. The thermodynamic limit extrapolated value for the critical temperature is obtained from
\begin{equation}
T_L=T_c+aL^{-1/\nu},
\end{equation}
where $a$ is a non-universal constant. From above equation we obtain $Tc =6.3535(3)$, where the smallest lattice has not been considered.

\begin{figure}[htb]
\includegraphics[width = 8.5cm]{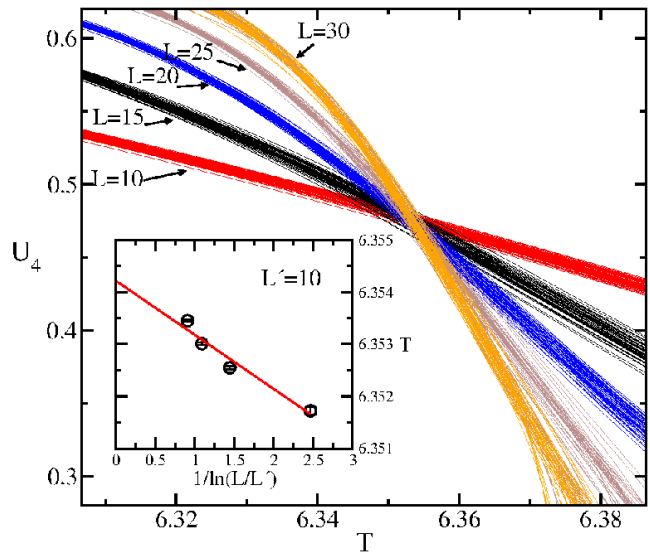}
\caption{(Color online) Fourth-order Binder cumulant $U_4$ as a function of temperature $T$, for $q=0$ and $50$ random samples for each lattice size ranging from $L = 10$ to $L=30$. The inset in the figure shows $T_L$ plotted versus the inverse of the logarithm of the scale factor $L/L^\prime$. The full line is the corresponding linear fit. The estimated temperature is $Tc =6.3542(2)$.}
\label{fig:2}
\end{figure}

Obtaining $Tc$ using the susceptibility maximum and the above equation requires, of course, the knowledge of the critical exponent $\nu$. From universality arguments one knows that this exponent is indeed dependent on the concentration \cite{murta}. However, another possibility to find $T_c$ is to use the finite-size scaling of the Binder cumulant, which for large enough lattices is independent of any exponent. We can then follow the variation of $U_4$ with $T$ for various system sizes and then locate the intersection of these curves for each sample, and finally calculating the mean and standard deviation of all intersections points. Fig. \ref{fig:2} depicts the results for $q=0$ (in this case we do not have dilution, only different MC runs. The results for other values of $q$ are rather similar, with a larger dispersion as we increase $q$ as noticed in the susceptibility behavior). All the 50 samples results are shown, for each lattice, in a narrow range of temperatures. 

The data have been obtained from histograms at different temperatures for each sample. We also opted here to compare the crossings of $U_4$ for two different lattice sizes $L$ and $L'$ of each sample. The statistics is now obtained from $50\times 50$ data. It is also clear from Fig. \ref{fig:2} that due to the presence of residual corrections to finite size scaling, one still needs to extrapolate the results by doing the limit $1/\ln(L/L') \rightarrow 0 $, as has been discussed in Ref. \cite{Loison1999}. This procedure is shown in the inset of Fig. \ref{fig:2} for $L^\prime=10$. From there we obtain $Tc =6.3542(2)$. This critical temperature for $q=0$  agrees very well with the previous one obtained from the susceptibility, and also with the high-temperature series expansion estimate by Butera and Comi $Tc=6.35435(3)$ \cite{Butera2000} and with other Monte Carlo results $Tc=6.35441(5)$ \cite{Lundow2009}, and  $Tc=6.3544(6)$ \cite{Diaz2012}. This is a quite nice numerical way for us to test the implementations that have been done in the computer codes.

In what follows, the transition temperature has been estimated using this process of cumulant crossings to avoid any use of critical exponents. In addition, we have resorted to Java and Python programming language with the use of some public libraries with NumPy and SciPy \cite{scipy}. It was then possible to link FORTRAN processing speed and automate the generation of the phase diagram in a simpler and more parallel way. This made possible to find a good adjustment of the experimental data, even using the present extensive MC simulations. A good adjustment was then obtained with the parameters $\zeta = 60$, $A = 1.25$ and $B = 0.75$.

\begin{figure}[htb]
\includegraphics[width = 8.5cm]{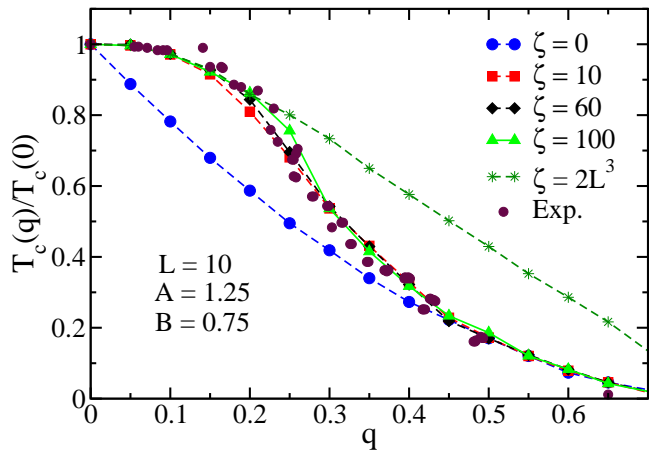}
\caption{(Color online) Phase diagram in the reduced temperature $T_c(q)/T_c(0)$ versus concentration $q$ plane for different values of the cluster size $\zeta$ for the lattice size $L=10$. As a matter of comparison, the experimental data coming from several references are also shown. The error bars are smaller than the symbol sizes and the lines are just a guide to the eyes.}
\label{fig:3}
\end{figure}

Before discussing the phase diagram itself, it is illustative to see the effect of the size of the clusters inducing the superexchange interaction. For this purpose, we have in Fig. \ref{fig:3} the reduced transition temperature $T_c(q)/Tc(0)$ as a function of $q$, for $A=1.25$, $B=0.75$ and different $\zeta$ values, for the lattice size $L=10$ (it is also included in this figure, as a matter of comparison, experimental data from references  \cite{Yelsukov1992} and  \cite{Stein2007}). We can observe from this figure that the behavior of the phase diagram is indeed strongly dependent on the cluster size parameter. When $\zeta= 0$, which represents the absence of superexchange interaction, a good fit is only observed for $q> 0.45$, with a finite and negative slope at $q=0$, contrary to experimental data. On the other hand, for $\zeta= 2L^3$ we have that the superexchange interaction is present in all Al cluster sizes, and a good fit is now obtained only for $q <0.2$. However, for $\zeta$  in range of $10$ to $100$, not only are the curves close together but also give a better fit to the whole experimental range of concentrations. A more detailed analysis of the data show that $\zeta= 60$ provides the best fit.

\begin{figure}[htb]
\includegraphics[width = 8.5cm]{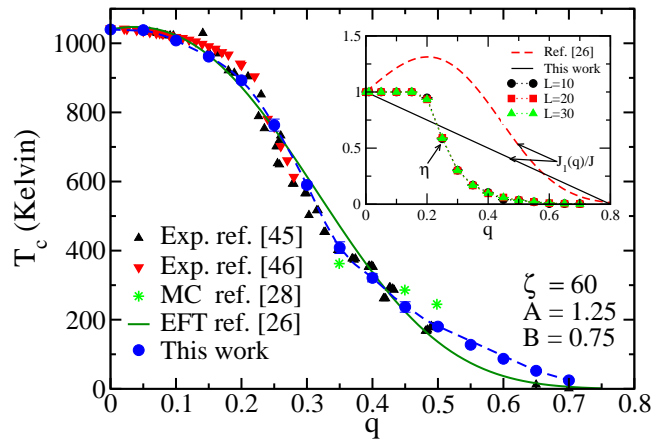}
\caption{(Color online) Phase diagram with the temperature $T_c$, now in Kelvin, as a function of Al concentration $q$ for our best fit parameters. Triangle up and triangle down are, respectively, from references  \cite{Yelsukov1992} and  \cite{Stein2007}. Stars come from MC simulations of reference  \cite{Ghosh2015} only in the concentration range $0.35 < q <0.5$. The solid line is the theoretical fit using the effective field theory from reference  \cite{Freitas2013}. Solid circles are the present Monte Carlo simulation results, where the dashed line is their spline univariate interpolation. The inset shows the behavior of the reduced nearest-neighbor interaction $J_1(q)/J$ as a function of $q$ according to Ref.  \cite{Freitas2013} (dashed line) and the present results (full line). 
It is also shown, in this inset, the Monte Carlo results of the ratio $\eta$,
defined in Eq. (\ref{eta}), for the largest lattices.
In this case, the dotted lines are just guide to the eyes and the error bars are smaller than the symbol 
sizes.}
\label{fig:4}
\end{figure}

Fig. \ref{fig:4} exhibits the transition temperature $T_c$, now given in Kelvin units, as a 
function of Al concentration $q$ for the best fit with $A=1.25$, $B=0.75$ and $\zeta=60$, in 
comparison with the experimental data \cite{Yelsukov1992, Stein2007}. The Monte Carlo results
have been obtained from finite-scaling extrapolation with $L=10,~15,~20,~25,$ and 30 for $q=0,~0.3$
and 0.7, while for the other concentrations only the cumulant crossings with $L=10$ and $L=20$ have
been considered. It is also shown in that figure the theoretical fit 
obtained by Freitas et al. using an effective field approach  \cite{Freitas2013} and Monte 
Carlo results by Ghosh et al. Ref. \cite{Ghosh2015}. One can clearly see that, apart from 
the tail for $q>0.5$, the present results are indeed in better agreement than those from 
effective field theory  \cite{Freitas2013}. In order to have a more quantitative comparison 
between both fits, we have also calculated the root mean square error (RMSE), the coefficient 
of determination, that is the square of the Pearson's product-moment correlation coefficient 
($r^2$),  and  Willmott index ($d$)  \cite{Willmott1985}. Since there are fewer simulation 
points than experimental data, we have calculate a spline univariate interpolation 
 \cite{spline} of the simulation data using python's SciPy library  \cite{scipy}. The 
interpolation is shown by the dashed line in Fig. \ref{fig:4}. The obtained results are 
summarized in Table I. The lower the RMSE value and the closer to $1$ are $r^2$ and 
$d$, the closer the model is to the experimental data (see, for instance, Ref.  \cite{Legates1999} 
and references therein). In this case, we observe the results obtained with the model proposed 
in this work is better than the model used by Freitas et al.  \cite{Freitas2013}. We did not 
make any finer comparison to the results by Ghosh et al.  \cite{Ghosh2015} because they only 
studied one narrow region of the phase diagram and, at the same time, it is visually clear that
the present transition curve is definitively closer to the experimental data than the results from \cite{Ghosh2015}. 

\begin{table}
\label{tab}
\begin{center}
\caption{The root mean square error (RMSE), the coefficient of determination ($r^2$),  and  Willmott index ($d$) obtained in this work and those calculated from the results of Ref. \cite{Freitas2013}.}
\begin{tabular}{l|ll}
		& { ~~This work~~}	& {~~~Ref. \cite{Freitas2013}} \\ \hline
 ~~RMSE~~	&~~~~ $3.5282$  &~~~ $4.3275$\\
 $~~~~~r^2$ 	&~~~~ $0.9949$	&~~~ $0.9924$ \\
 $~~~~~d$   	&~~~~ $0.9972$  &~~~ $0.9959$ \\ \hline
\end{tabular}
\end{center}
\end{table}

Besides the better agreement of the present approach over the model without any induced next-nearest-neighbor interactions of ref.   \cite{Freitas2013}, we can also notice the rather strange behavior that the nearest-neighbor interaction must have, as a function of the Al concentration $q$, when no extra couplings are considered in the theoretical model. This is shown in the inset of Fig. \ref{fig:4}. In this case, in order to keep the transition temperature almost constant with dilution, without the presence of a superexchange interaction, the nearest-neighbor coupling $J_1(q)$ has to increase as $q$ increases, in order to balance the broken bonds caused by the diluted atoms. And it indeed increases about 30\% for concentrations up to $q\sim 0.22$. However, in view of the calculations from first principles  \cite{Ghosh2015} this is not expected, at least in this small range of Al concentrations. This means that a model that takes into account the influence of the size of Al clusters inducing extra interactions in these alloys should be more appropriate to a direct connection to the thermodynamics of the magnetic properties of Fe-Al compounds. 

Concerning the adjustments of the theoretical parameters $A$, $B$ and $\zeta$, one should mention that,
apart from $\zeta$, which has been obtained in a process depicted in Fig. \ref{fig:3}, the first two
of them can in fact be roughly estimated, since they affect different regions of the phase diagram, making it easier to fit the critical transition line. 
For instance, it is noticed that the parameter $B$ mostly influences the initial anomalous region of the diagram 
($ q <0.2 $). In this region, the Al atoms are more likely to be isolated from each other, so each one 
breaks $8$ nearest-neighbor exchange interactions $J_1$ and, at the same time, can induce up to most $12$ new 
superexchange interactions $J_2$. In order for the dilution effect, in this initial region of the phase 
diagram, be as minimal as possible, the value of $B$ should be at least $2/3$, keeping in this way the 
balance of the ferromagnetic interactions. However, as clusters with more than one Al atom can indeed be 
formed in the alloys, together with the fact that the nearest-neighbor interaction decreases with 
Al concentration, the best value for the fits should be a little higher, and in the present case it is 
found $B=0.75$. On the other hand, the $A$ parameter is mainly used for fits of the 
diagram in the higher Al concentration range. Its value has been chosen so that there is no 
antiferromagnetic nearest-neighbor interactions, since neutron scattering data for $q>0.25$ showed no 
indication of any antiferromagnetic order \cite{Shull1976}. This means that $J_1(q)$ must be positive 
for concentrations up to $q\sim 0.8$, which is just above the critical value $q_c=0.7540385(10)$. So, the choice $A=1.25$. A similar idea has been already used 
in a previous three-dimensional diluted ferromagnetic XY model \cite{SantosFilho2011} to adjust the experimental data of the insulating pentacoordinate iron(III) molecular ferromagnet Fe[DSC]$_2$Cl. 

Finally, the present numerical values for the theoretical parameters are comparable to the previous
ones using analytical approaches, at least concerning the nearest-neighbor interaction $J_1(q)$.
For instance, as the Boltzmann constant $k_B = 8 . 617 \times 
10^5$ eV/K and the experimental result of the critical temperature of the pure Fe system is 
$T_c = 1040$ K, we obtain $J = 0.014$ eV, a value that is in fully agreement with  
$J = 0.013$ eV from reference \cite{Plascak2000} and $J = 0.014$ eV from reference \cite{dias1}. 
This agreement comes from the fact that the transition temperature coming from the employed
approximations in those works are comparable to the present Monte Carlo result for the bcc lattice.
The value of $A=1.25$ is also comparable to $A=0.85$ and $A=0.95$ obtained, respectively, in references
\cite{Plascak2000} and \cite{dias1}. Notice also that even $\zeta=60$ is of the order of the size of the gold clusters ($\sim 100$) with molecular-metal behavior \cite{Mustalahti2015}.

The situation is, however, different concerning the superexchange
interaction. In the previous works $J_2$, assumed to depend on the concentration $q$, 
has a kind of oscillatory behavior, becoming still negative for some range of $q$,
which is not expected for this system. In the present approach, $J_2$ 
does not depend on Al concentration, but is instead influenced by the number of clusters that
can induce the superexchange interactions. This can be better seen by tracking the
quantity
\begin{equation}
\eta={{N_s}\over{N_t}},
\label{eta}
\end{equation}
where $N_s$ is the average effective number of superexchange interactions induced by 
clusters smaller than $\zeta$ and $N_t$ is the average of the total number of 
next-nearest-neighbor Fe atoms surrounding all clusters of the sample. The corresponding 
behavior of $\eta$ for the fitted parameter $\zeta=60$ is also depicted in the inset of 
Fig. \ref{fig:4}. One can clearly see that for low values of the concentrations one has $N_s\sim N_t$,
due to the fact that here larger clusters are seldomly formed.
But, as the concentration increases, larger clusters are formed and accordingly decreasing the number of superexchange interactions.

\section{Concluding remarks}

We have used a site diluted spin-1/2 Ising model, with nearest- and nex-nearest-neighbor interactions, 
to describe the magnetic phase diagram of Fe-Al alloys in the bcc lattice. We have employed Monte Carlo 
simulations based on a hybrid algorithm consisting of one single-spin Metropolis move and one single-
cluster Wolff algorithm. In some regions of the phase diagram we have also used single histogram 
techniques allied with finite-size scaling arguments. With the assumption that only aluminum present in 
clusters composed of less than 60 Al atoms induce a superexchange interaction among next-nearest-neighbor Fe atoms surrounding the cluster, a good description of the phase diagram could be obtained, including the anomalous region for small Al concentrations.

We have used here Monte Carlo simulations because it seems to be the most suitable approach to properly take into 
account the Al cluster size effects. Such cluster size properties have not been considered 
previously and better fits than those from Refs.  \cite{Perez1986,Plascak2000,sala,mota,osorio,Contreras2005,aguirre,lara,dias1,dias2,Diaz2012,Freitas2013,freitas} have been achieved. In addition, Monte Carlo simulations provide not only more precise values for the transition temperatures, but also can capture the real physics of the cluster size properties. For this 
reason, we believe that the fitted theoretical parameters should also be more accurate. In addition,
the behavior of the exchange and superexchange interactions with the Al concentration is physically
more plausible than those earlier reported for these alloys.

We have not estimated the errors of the fitting parameters in the usual way. We have, however, done some extra simulations just to test the order of sensitiveness the fits exhibit by changing the values
of them. For example, the quality of the fit has no significant changes by considering $\zeta =55$ 
or $60$, $A=0.7$ or $0.8$ and $B=1.3$ or $1.2$.
	
As a final remark, we have not considered the possibility of having a nearest-neighbor exchange interaction that increases as $q$ increases, as pointed out recently in reference \cite{Ghosh2015} for $0.35< q < 0.5$. We believe that such oscillatory behavior will not produce significant changes either in the anomalous region or even out of it. In fact, the corresponding Monte Carlo results shown by the stars in Fig. \ref{fig:4} do not exhibit a good agreement neither with the experimental data nor with the theoretical model predictions. 

\begin{acknowledgments}
The authors would like to thank A. S. Freitas for the availability of the theoretical data obtained using the effective field theory and M. F. Cavalcante for fruitful discussions. Financial support from the Brazilian agencies CNPq and FAPEMIG are also gratefully acknowledged. 
\end{acknowledgments}

\end{document}